\documentclass[a4paper, amsfonts, amssymb, amsmath, reprint,reprintnumbers, showkeys, nofootinbib, twoside, floatfix]{revtex4-1}
\usepackage{float}
\usepackage{dsfont}
\usepackage[english]{babel}
\usepackage[utf8]{inputenc}
\usepackage[colorinlistoftodos, color=green!40, prependcaption]{todonotes}
\usepackage[pdftex, pdftitle={Article}, pdfauthor={Author}]{hyperref} 
\usepackage{graphicx}
\usepackage{epsfig,rotate,latexsym,amsmath,mathrsfs,amssymb}
\usepackage{xcolor,hyperref}
\usepackage[arrowdel]{physics}
\hypersetup{final}
\def \be{\begin{equation}}
\def \ee{\end{equation}}

\begin{document}

\title{Secondary Hadron--Nucleus Collisions of Short-Lived Hadrons in 
Ultra-Relativistic Fixed-Target Heavy-Ion Interactions}

\author{Sanatan Digal}
\email{digal@imsc.res.in}
\affiliation{Institute of Mathematical Sciences, Chennai 600113, India}
\affiliation{Homi Bhabha National Institute, Training School Complex, 
Anushaktinagar, Mumbai 400094, India}
\author{P. S. Saumia}
\email{saumia@gmail.com}
\affiliation{LifeSignals India Pvt Ltd, Bangaluru 560034, India}
\author{Ajit M. Srivastava}
\email{ajit@iopb.res.in}
\affiliation{Institute of Physics, Bhubaneswar 751005, India}
	
\begin{abstract}
	Ultra-relativistic heavy nuclei traversing a solid target undergo 
successive nuclear encounters separated by atomic lattice spacings. At 
sufficiently high beam energies, Lorentz contraction reduces the proper time 
between collisions to $\mathcal{O}(10^4)$~fm$/c$ in the center-of-mass frame 
of the first interaction. We then consider the fragmentation region of this 
first collision, and show that short-lived hadrons produced in this region,
with additional Lorentz boost, can reach the next nucleus before decaying.  
We show that this geometry enables secondary hadron--nucleus collisions 
involving species that cannot be realized as conventional secondary beams 
or in subsequent hadron--nucleus interactions in cosmic-ray cascades. For a 
$2.76$ TeV-per-nucleon Pb beam incident on a solid Pb lattice, we determine 
which forward-produced hadrons can survive to a second interaction, estimate 
their collision probabilities, and analyze potential observable consequences.  
In particular, we identify some representative hadrons whose proper lifetimes 
are of order $10^3$ fm/c, e.g. specific mesons ($\eta^\prime$) and heavy-flavor 
resonances ($J/\psi, D^*(2010)$), as projectile species that become accessible 
through this collision space-time geometry. At substantially higher beam 
energies (for example, with 10 TeV per-nucleon Pb beam), the survival 
probabilities are significantly enhanced. This can make even very short lived 
hadrons with life times of few tens fm ( $\Xi$(1530), $\omega$(782), $\phi$(1020)) 
available for this secondary hadron-nucleus collision, providing an additional 
motivation for future ultra-relativistic fixed-target heavy-ion experiments.
		\end{abstract}
\maketitle	

\section{Introduction}

\par Hadron--nucleus ($h$--$A$) and nucleus--nucleus ($A$--$A$) collisions provide 
fundamental probes of strong-interaction physics ~\cite{Busza:1975te,Koplik:1975ni,
rafelski,Bjorken:1982qr,armesto_shadowing,heinz2013,  schukraft,busza2018, alice}. 
These reactions enable the study of phenomena ranging from hadronic multiple scattering 
in cold nuclear matter to collective behavior, strangeness production, and deconfinement 
in relativistic heavy-ion collisions~\cite{rafelski,Gazdzicki:2010iv,heinz2013,busza2018}. 
Over several decades, extensive experimental and theoretical efforts have established a 
detailed understanding of proton-- and pion--nucleus interactions, including nuclear 
shadowing, energy loss, baryon stopping, and in-medium hadron production, providing 
essential baselines for interpreting heavy-ion data~\cite{schukraft,alice,urqmd,jam}.

Experimental studies of hadron--nucleus collisions have progressively
extended beyond proton- and pion-induced reactions to include strange
hadrons. Secondary kaon and hyperon beams at the CERN SPS have enabled
detailed studies of hyperon production and hyperon--nucleus interactions
~\cite{Bourquin:1984kcf,Aleksandrov:1997kj,Gal:2016boi}. 
In particular, the WA89 experiment used secondary $\Sigma^-$,
$\pi^-$, and neutron beams to investigate strange-particle production in
interactions with nuclear targets~\cite{MIA1}. The same experimental program
also studied the production of charmed hadrons in $\Sigma^-$--nucleus
interactions~\cite{MIA2}. Secondary charged kaon beams have also been developed
at J-PARC for studies of hypernuclei and double-strangeness systems ~\cite{Agari:2012kid}.
Dedicated hypernuclear programs, such as with PANDA at FAIR, have further
explored the production of unstable hyperons in a primary reaction,
followed by their transport to and interaction or capture in a secondary
target ~\cite{PANDA:2009yku,PANDA:2020zwv,Gaitanos:2011fy}. Together, these 
studies provide an important experimental
foundation for strange hadron--nucleus interactions.

These approaches, however, rely on particles that survive long enough to be
produced, transported and focused as secondary beams, or, in the case of
PANDA-like configurations, transported to a secondary target. Consequently,
hadrons with proper lifetimes of only a few hundred to a few thousand fm/$c$,
including many excited strange hadrons and heavy-flavor resonances~\cite{pdg}, 
cannot realistically be prepared and employed as conventional secondary 
projectile beams. While interactions of such short-lived hadrons with nuclear 
matter have been investigated indirectly through their production and propagation 
in $p$--$A$ and $A$--$A$ collisions~\cite{rapp_heavy_flavor,andronic_heavy}, 
their use as projectile beams in direct hadron--nucleus collision experiments has 
remained largely inaccessible. Direct hadron--nucleus collision experiments involving 
such projectiles therefore remain largely unexplored, and extending experimental 
access to this short-lifetime regime is the primary motivation for the present work.

Cosmic-ray studies provide a complementary probe of hadron--nucleus interactions at energies extending well beyond those available at terrestrial accelerators. Extensive air-shower measurements have yielded valuable information on forward particle production and proton--air cross sections, and have played a central role in the development and validation of hadronic interaction models~\cite{gaisser1990,engel2011,heck1998,pierog2015,ostapchenko2011,Riehn:2019jet,enberg2008,abreu2012}. However, the interaction geometry differs fundamentally from that in a solid target. Successive interactions occur only after propagation over macroscopic distances through the atmosphere. Consequently, unstable hadrons produced in one interaction can serve as projectiles in subsequent hadron--nucleus collisions only if their lifetimes are sufficiently long. Hadrons with proper lifetimes of order a few thousand fm/$c$ or shorter decay well before encountering another atmospheric nucleus and therefore contribute to air-shower development only through their decay products rather than as incident projectiles~\cite{enberg2008}. Cosmic-ray interactions therefore do not provide direct access to hadron--nucleus collisions in this short-lifetime regime.

In this work we analyze ultra-relativistic fixed-target heavy-ion interactions in a solid lattice as a novel environment for secondary hadron--nucleus scattering. When an energetic Pb nucleus traverses (a reasonably well aligned) solid Pb target, successive lattice nuclei along the beam direction are separated by $4.95\,\text{\AA}$. In the center-of-mass (CMS) frame of the first Pb--Pb collision, Lorentz contraction reduces this separation to about $10^4$ fm. This space--time geometry allows short-lived hadrons produced in the forward fragmentation region, with very large Lorentz boosts in the CMS frame, to survive long enough to encounter a subsequent target nucleus, thereby extending hadron--nucleus collisions into the short-lifetime regime identified above. We determine the kinematic conditions under which such processes become possible, evaluate the lifetime constraints on the produced hadrons, and estimate the corresponding secondary-collision probabilities. We briefly discuss the potential implications of these interactions and their potential experimental manifestations.

\section{COLLISION GEOMETRY AND KINEMATICS}

We consider a Pb nucleus, with beam energy, $E_{\rm beam}=2.76$ TeV per nucleon 
colliding with a target lattice of Pb atoms. The lattice is considered to be cubic and spacing between nearest neighbour(NN) Pb atoms, $d_{\rm lab} \simeq 4.95\,\text{\AA} = 4.95\times10^5\ \text{fm}$. Further we assume that the projectile nucleus is reasonably well aligned with a line of Pb atoms in the lattice. Although the present setup is fixed-target, the first interaction defines a well-defined Pb–Pb center-of-mass frame (CMS frame), which we denote as $S_0$; the dynamics can be analyzed in both the laboratory (Lab) and this CMS frame, with the latter providing the most transparent description for subsequent space–time analysis.

The nucleon--nucleon center-of-mass energy for $E_{\rm beam}=2.76$ TeV is,
\begin{equation}
\sqrt{s_{NN}} = \sqrt{2m_N E_{\rm beam}} \approx 72\,\text{GeV}.
\end{equation}
The associated Lorentz factor ($\gamma_0$) of each nucleus in the CMS frame $S_0$ is:
\begin{equation}
\gamma_0 = \frac{\sqrt{s_{NN}}}{2m_N} \approx 38.4,
\end{equation}
and their rapidities are given by,
\begin{equation}
	y_0 = \pm \cosh^{-1}(\gamma_0)  \simeq \pm 4.34.
\end{equation}

After the initial Pb–Pb collision, a quark–gluon plasma (QGP) forms and 
expands, eventually hadronizing into a hadron gas that continues to expand. 
Some of these hadrons—particularly those in the fragmentation region—undergo 
secondary collisions with Pb nuclei in the target lattice. In the lab frame, 
in the target lattice the nearest Pb nucleus is 
$d_{\rm lab}\simeq 4.95\text{\AA}$ away. In the CMS frame, $S_0$, of the 
first Pb-Pb collision, $d_{\rm lab}$ is Lorentz contracted to a value $d_0$
given by,

\begin{equation}
	d_0 = \frac{d_{\rm lab}}{\gamma_0} 
\approx 1.3\times10^4\ \text{fm}.
\end{equation} 

For some of the exotics in the central rapidity, with short life time, this 
separation could be too large. On the other hand, when they are in the forward 
fragmentation region, their life time gets dilated enabling them to survive 
until they reach the next Pb-nucleus in the target lattice. In the following, we 
consider these hadrons and evaluate their survival probability up to the 
subsequent Pb nucleus, as well as their collision probability.
For the propagation of these secondary hadrons to the next target nucleus, it is 
useful to work in the laboratory frame, in which the target lattice is at rest. However,
we retain the center-of-mass frame $S_0$, which gives a clear picture of the first 
Pb--Pb collision and kinematics of the fragmentation-region.

The physics of the fireball in the fragmentation region is very well discussed in
the literature \cite{frag}. The peak baryon density in the fragmentation 
region in the immediate aftermath of the first collision may reach 
$n_B\approx (3-8)n_0$, where it undergoes subsequent hydrodynamical evolution as a 
baryon rich fireball. Evidence from baryon stopping at RHIC and SPS energies 
suggests rapidity loss of $\Delta y\approx 1.45-2.45$ 
\cite{brahms_netproton, brahms_arsene}. In the following we 
 present result for two values of rapidity loss,
$\Delta y = 1.45$, and 2.45\cite{brahms_netproton}. 

For a hadron with initial rapidity $y_0 = 4.34$, and $\Delta y = 1.45$, 
the corresponding velocity and Lorentz factor in frame $S_0$ 
are $v_{\rm frag} = 0.9938$ and $\gamma_{\rm frag} = 9.0$, respectively. 
To account for the hydrodynamic expansion of the fragmentation region, we 
introduce an additional boost to the leading hadrons, taken to be of the order 
of the speed of sound, $v_{\rm exp} = 1/\sqrt{3}$, which results in  
$v_{\rm frag} = 0.9984$ and 
$\gamma_{\rm frag} = 17.4$ (in frame $S^0$).  
For $\Delta y = 2.45$, 
the corresponding values are $v_{\rm frag} = 0.9878$ and 
$\gamma_{\rm frag} = 6.43$. In the laboratory frame, 
the Lorentz factors 
increase to $\gamma^L_{\rm frag} = 1333$  for $\Delta y = 1.45$ and 
$\gamma^L_{\rm frag} = 491$ for $\Delta y = 2.45$.
These laboratory-frame Lorentz factors determine the Lorentz-dilated lifetime of 
the hadrons in the laboratory frame and hence enter directly into the 
calculation of their survival probability to the neighboring Pb nucleus.

\section{SURVIVAL PROBABILITY OF SHORT-LIVED
HADRONS}

In the laboratory frame, the hadrons are ultra-relativistic.
Consequently, their lifetime is highly time-dilated, 
$\tau = \gamma^L_{\rm frag}\,\tau_0$, where $\tau_0$ denotes 
the proper lifetime. 
The time required for a hadron to reach the next Pb nucleus is approximately 
$d_{\rm lab}/v^L_{\rm frag}$ where $v^L_{\rm frag}$ is the velocity of
hadron corresponding to its Lorentz boost factor 
$\gamma^L_{\rm frag}$. Combining these results, the survival probability of 
secondary hadrons to reach the next Pb nucleus, in the target lattice, is given by
\begin{equation}
	P_{\rm surv} = \exp\!\left(-\frac{d_{\rm lab}}{\gamma^L_{\rm frag}\,\tau_0\,v_{\rm frag}}\right).
\end{equation}
With $d_{\rm lab} =  4.95\text{\AA}$, and $v_{\rm frag} \simeq c$, we write
below expressions of two probabilities using the two values   
$\gamma^L_{\rm frag} = 1333$, and $\gamma^L_{\rm frag} =$ 491 ,
corresponding to $\Delta y = 1.45$ and 
$\Delta y = 2.45$, respectively.

\begin{equation}
P_{\rm surv}(\Delta y = 1.45) = 
\exp\!\left(-\frac{371.34}{\tau_0(\rm fm)}\right), ~~{\rm and},
\end{equation}

\begin{equation}
P_{\rm surv}(\Delta y = 2.45) 
= \exp\!\left(-\frac{1008.15}{\tau_0(\rm fm)}\right).~~~~~
\end{equation}

These expressions show that hadrons with proper lifetimes of order a few hundred to a few thousand fm/$c$ can survive long enough to undergo secondary hadron--nucleus collisions with the neighboring Pb nucleus, even for the larger rapidity loss, $\Delta y = 2.45$. Hadrons with substantially longer lifetimes, such as $K_S^0$, $\Lambda$, $\Sigma$, and $\Omega$, can readily survive over these distances and have already been investigated using conventional secondary beams and hypernuclear experiments. They are therefore not the primary focus of the present work. Instead, we concentrate on the short-lifetime regime that has remained inaccessible to direct hadron--nucleus collision experiments. Representative examples of a few such hadrons, together with their survival probabilities for the two values of rapidity loss considered here, are listed in Table~\ref{tab:all_weak}. Among these, the survival probabilities are substantial for hadrons with lifetimes of order $10^3$ fm/$c$, while they become strongly suppressed for much shorter-lived resonances such as $\Xi(1530)$, $\omega$, and even for $\phi$ with $\Delta y=2.45$. For the latter, the survival probability remains potentially observable only for smaller rapidity loss, e.g. $\Delta y=1.45$.

\begin{table}[htb]
\centering
\begin{tabular}{lccc}
\hline\hline
Hadron & $\tau_0$ (fm/$c$) & $P_{\rm surv}(\Delta y=1.45)$ & $P_{\rm surv}(\Delta y=2.45)$ \\
\hline
$\Xi(1530)$ & $\approx 20$ & 8.68$\times 10^{-9}$ & $\approx 0$ \\
$\omega(782)$ & $\approx 23$ & 9.78$\times 10^{-8}$ & $\approx 0$ \\
$\phi(1020)$ & $\approx 45$ & 2.61$\times 10^{-4}$ & $2 \times 10^{-10}$ \\
$\eta^\prime$ & $\approx 1000$ & 0.69 & 0.364 \\
$J/\psi$ & $\approx$ 2200 & 0.84 & 0.63 \\
$D^*(2010)^\pm$ & $\approx$ 2500 & 0.86 & 0.67 \\
$\eta$ & $\approx 10^5$ & $\approx 1$ & 0.99 \\
\hline\hline
\end{tabular}
\caption{Proper lifetimes (from the Particle Data Group~\cite{pdg}) and corresponding survival probabilities for representative short-lived hadrons produced in fixed-target Pb--Pb collisions at $E_{\rm beam}=2.76$~TeV per nucleon. The survival probabilities are shown for two representative values of the rapidity loss, $\Delta y=1.45$ and $2.45$.}
\label{tab:all_weak}
\end{table}

Table~\ref{tab:all_weak} illustrates that with $E_{\rm Lab}=2.76$~TeV, the transition between 
inaccessible and 
accessible projectile species occurs for proper lifetimes of order of a thousand fm/$c$ 
for $\Delta y=2.45$, and few hundred fm/$c$ for $\Delta y=1.45$. These lifetime scales naturally emerge from the combined effects of the Lorentz-contracted spacing of successive target nuclei and the Lorentz dilation experienced by forward-produced hadrons. At lower beam energies the accessible lifetime window shifts toward 
much longer-lived particles. For example, for $E_{\rm beam}=158$ GeV, the survival probability of a hadron with $\tau_0=10^3$ fm/$c$ is only $1.53\times10^{-3}$ and $2.20\times10^{-8}$ for $\Delta y=1.45$ and $\Delta y=2.45$, respectively. These results suggest that secondary hadron--nucleus collisions involving such short-lived resonances would have been strongly suppressed at CERN-SPS energies, while becoming increasingly accessible at higher fixed-target beam energies. At substantially higher beam energies (for example, with 10 TeV per-nucleon Pb beam), the survival probabilities are significantly enhanced. For example, survival probabilities of very short lived hadrons in Table I with life times of few tens fm ($\Xi$(1530), $\omega$(782), $\phi$(1020)) are increased (using $\Delta y$ = 1.45) to $\approx 10^{-3} - 10^{-1}$. This can make even these hadrons available for this secondary hadron-nucleus collision. This may be considered as providing an additional motivation for future ultra-relativistic fixed-target heavy-ion experiments with higher energies.

\section{GEOMETRIC AND THERMAL OVERLAP FACTORS}

 Now we consider additional geometric factors in determining the final 
probability of these {\it secondary}  hadron-nucleus collisions. 
At times $t \gg 10$ fm/$c$ the system is expected to have ceased hydrodynamic 
evolution and is expected to be in a free-streaming regime. Hence, the 
transverse extent of the initial fireball from the first Pb-Pb collision to 
time $t_c$, when the leading hadrons reach the next Pb nucleus, is mostly 
governed by ballistic expansion of the bulk hadronic debris.
 We therefore estimate the transverse
size using an effective transverse expansion velocity
$v_\perp\simeq 0.5c$. For the $\Delta y=2.45$ case, the proper time
corresponding to propagation to the neighboring nucleus is
\begin{equation}
\tau \simeq \frac{d_{\rm lab}}{c\,\gamma_{\rm frag}^{L}}
\simeq 1008~{\rm fm}/c ,
\end{equation}
giving
\begin{equation}
R_{\rm trnsvrs}\simeq v_\perp \tau
\simeq 504~{\rm fm}.
\end{equation}
This gives the geometric overlap factor with the transverse size of the
neighboring Pb nucleus, with transverse size $R_{\rm Pb}\simeq 7$ fm, as

\begin{equation}
f_{\rm geom} \approx \left(\frac{R_{\rm Pb}}{R_{trnsvrs}}\right)^2
\approx 1.92\times 10^{-4}.
\label{geo}
\end{equation}
Note that the forward-going hadrons from the fragmentation region will occupy a finite 
transverse extent as they propagate toward the next nucleus, and this distribution can be 
viewed geometrically as defining a cone whose apex is at the primary collision point. For a 
secondary collision to occur, the neighboring nucleus must lie within this cone. The full cone 
angle is approximately
\begin{equation}
\theta_{\rm cone} \simeq \frac{2R_{\rm trnsvrs}}{d_{\rm lab}}={1 \over \gamma^{L}_{\rm frag}}.
\end{equation}
For $\gamma^{L}_{\rm frag}=491$ with ($R_{\rm trnsvrs}\simeq 504$ fm and $d_{\rm lab}\simeq 4.95\AA$), we get 
$\theta_{\rm cone}\simeq0.12^\circ$. Though, the cone angle is small, it is certainly possible 
to align the beam so that the next nucleus falls within this cone.

The above geometrical estimate, Eq.\ref{geo}, assumes that the target nuclei remain at 
their equilibrium lattice positions. In a real solid, however, the atoms undergo thermal vibrations 
about these equilibrium positions governed by the Debye temperature \cite{thermalreview}. These fluctuations can be important because the transverse 
size of the target nucleus is much smaller than the typical transverse displacement of the 
lattice atom over the relevant length scale. We therefore include the thermal motion of the 
target nucleus in estimating the overlap probability.

We characterize the thermal displacement of the target atom by the one-dimensional rms 
amplitude $u_1$, which can be obtained from the tabulated Debye--Waller $B$ factors~\cite{Bfactor}. 
Assuming independent fluctuations in the two transverse directions, the corresponding rms 
transverse displacement is $\sqrt{2}u_1$. The additional overlap factor associated with the 
thermal motion can then be estimated as
\begin{equation}
f_{\rm thermal} \simeq
\left(\frac{R_{\rm trnsvrs}}{\sqrt{2}u_1}\right)^2.
\end{equation}

The total overlap factor is consequently obtained by combining the geometrical overlap with the 
thermal-motion factor,
\begin{eqnarray}
f_{\rm total}
&=& f_{\rm geom}\times f_{\rm thermal} \\
&=& \left(\frac{R_{\rm Pb}}{\sqrt{2}u_1}\right)^2 .
\label{ftl}
\end{eqnarray}
An interesting feature of this estimate is that the transverse size
$R_{\rm trnsvrs}$ cancels from the final overlap factor. Thus, within
the present approximation, the beam energy and rapidity loss affect the
survival probability through $\gamma_{\rm frag}^{L}$, while the final
overlap factor is controlled by the nuclear radius and the thermal
displacement of the target nucleus.

For Pb at room temperature, the tabulated Debye--Waller factors give
$\sqrt{2}u_1\simeq 0.21\text{\AA}=0.21\times10^5\text{fm}$~\cite{Bfactor}.
Using $R_{\rm Pb}\simeq 7\text{fm}$, we obtain
\begin{equation}
f_{\rm total}^{\rm Pb}
\simeq
\left(\frac{7\text{fm}}{0.21\times10^5\text{fm}}\right)^2
\simeq 1.1\times10^{-7}.
\end{equation}

The overlap can be enhanced by reducing the target temperature. For
example, at temperature $85\text{K}$ the thermal displacement of Pb is reduced by
approximately a factor of two~\cite{Bfactor}, resulting in an approximately
fourfold increase in $f_{\rm total}$. The factor will be even larger with decreasing  value of temperature, we have used values from the tables in \cite{Bfactor} where 85 K is the lowest value listed. A further improvement may be obtained by using a target material with a larger Debye temperature and hence smaller
thermal displacement. As an example, tungsten, with mass number $A=184$,
has a substantially larger Debye temperature than Pb,
$\Theta_W\simeq310\text{K}$ compared with $\Theta_{\rm Pb}\simeq88\text{K}$.
The corresponding tabulated values give $\sqrt{2}u_1\simeq0.06\text{\AA}$ for W at
room temperature and $\sqrt{2}u_1\simeq0.03\text{\AA}$ at $85\text{K}$~\cite{Bfactor}.
Taking $R_W\simeq6.7\text{fm}$, this gives
\begin{equation}
f_{\rm total}^{\rm W}
\simeq
\left(\frac{6.7\text{fm}}{0.03\times10^5\text{fm}}\right)^2
\simeq5.0\times10^{-6},
\end{equation}
which is significantly larger than the corresponding
Pb value. Further, in the case of Tungsten, the lattice spacing being $\approx 40\%$ smaller 
would enhance the $P_{\rm surv}$ listed in Table.I

\section{SECONDARY COLLISION PROBABILITY
AND OBSERVABLE CONSEQUENCES}

The probability per primary Pb--Pb event that a produced hadron of
species $h$ undergoes a secondary interaction with a lattice nucleus
can be estimated as
\[
P_{hA} \approx N_h \times P_{\rm surv} \times f_{\rm total},
\]
where $N_h$ is the average multiplicity of species $h$ in the forward
fragmentation region. Note that this quantity represents an event-level
interaction probability rather than a differential scattering cross section.
The interaction probability per hadron is then seen to be  
$\approx 10^{-6}$ for all hadrons except for $\Xi$, $\omega$, and $\phi$.
For $\phi$, the interaction probability (per $\phi$ particle produced) 
is $\approx 10^{-10}$ for lower value of rapidity loss, 
$\Delta y = 1.45$.

The estimates presented above suggest that primary hadrons produced in 
ultra-relativistic fixed-target heavy-ion collisions can potentially undergo secondary 
collisions with neighboring nuclei in the target lattice. 
In the lab frame, it will appear as the respective hadron colliding with neighboring nucleus with energy
$E_{hadron} = \gamma^L_{frag} m_{hadron}$ where $\gamma^L_{frag} = 1333$ and 491 for $\Delta y = 1.45$
and 2.45 respectively. (Thus, in the lab frame, $J/\psi$ produced in the primary interaction will collide 
with the neighboring Pb nucleus with $E_{J/\psi} = $ 4.1 TeV, and 1.5 TeV for the two values of rapidity 
loss $\Delta y$.) 

Although the probabilities are small, these processes extend hadron--nucleus scattering into a 
short-lifetime regime that is inaccessible using conventional secondary beams or cosmic-ray interactions. 
In particular, excited hyperons and selected heavy-flavor resonances with proper lifetimes of order 
$10^3$ fm/$c$ become potential projectile species for direct hadron--nucleus collisions. 
Observable signatures of these secondary interactions may be expected to be suppressed by the much 
larger particle multiplicity from the primary collision. To isolate these signals, focus has to be on 
large rapidity regions, in the forward fragmentation region, which may help to separate from the large
multiplicities of the central rapidity region. 

It will be important to see effects of varying 
centrality in our model. This will require much more detailed understanding of the evolution of
fragmentation region for different centralities. One interesting possibility may be to rotate
the target lattice so that the primary beam points in different crystallographic directions. Particle
distributions in the forward fragmentation region should change with the changes in the rotation angle
as the distance to the next nucleus along beam direction will change, with maximum effect of secondary
interactions being visible for the {\it aligned} direction with the shortest distance to the next nucleus.
Importantly,  similar distributions should be obtained on the two sides of this {\it aligned} direction. 
It is difficult at this stage to give any quantitative estimate of the effects of such rotation because our 
estimates of geometric overlap factor (Eq.~\ref{ftl}) are very crude. 

We have used simple estimates here with the main aim to draw attention to this interesting possibility.
Detailed estimates will be needed to explore the viability of this proposal. One needs realistic 
angular distributions of produced hadrons from the evolution of the baryon rich fragmentation region 
and account for fluctuations,  rescattering effects etc. A proper study of this will require a transport calculation
(e.g., using UrQMD).
At this stage our main aim is only to point out this interesting possibility of directly studying 
hadron--nucleus collisions involving such short-lived projectiles.  This also provides additional 
motivation for future investigations of in-medium hadron interactions and their implications for 
hypernuclear physics and in-medium hadron interactions involving  heavy flavors.

The particles produced in the collision of secondary hadrons with the
next Pb nucleus in the target lattice will affect the space time structure of the primary collisions. However, the multiplicity from the primary collision 
dominates the final state, thus the contribution of secondary interactions
to two-particle correlation observables such as HBT interferometry is
expected to be suppressed. Nevertheless, in specially selected kinematic windows
or geometries, small deviations from conventional source functions
cannot be excluded.

\section{SUMMARY AND CONCLUSIONS}

In conclusion, we have demonstrated that ultra-relativistic fixed-target 
heavy-ion collisions in a solid lattice enable secondary hadron--nucleus 
interactions involving hadrons that cannot serve as projectiles in cosmic 
rays or accelerator-based beams. These include excited hyperons and heavy-flavor resonances with proper lifetimes of order $10^3$ fm/c, which cannot be realized as conventional secondary beams. Our results therefore identify an experimentally unexplored lifetime window for hadron--nucleus collisions that could become accessible in ultra-relativistic fixed-target heavy-ion experiments, especially at higher energies
in future for hadrons of even shorter lifetimes. 

\begin{acknowledgments}
A. M. Srivastava acknowledges support from the Raja Ramanna Chair position (DAE, Government of India).
\end{acknowledgments}

\end{document}